
\overfullrule=0pt
\magnification=\magstep1
\baselineskip=3ex
\raggedbottom

\def\uprho{\raise1pt\hbox{$\rho$}}
\def\mfr#1/#2{\hbox{${{#1} \over {#2}}$}}
\def\upchi{\raise1pt\hbox{$\chi$}}
\def\ch{{\cal H}}
\def\bs{{\bf S}}
\def\R{{\bf R}}
\def\J{{\bf J}}
\def\1{{\bf 1}}
\def\Tr{{\rm Tr}}
\def\d{{\rm d}}
\headline={\hfill{\fiverm 13/Nov/93}}
\centerline{\bf THE HUBBARD MODEL:}
\centerline{\bf SOME RIGOROUS RESULTS AND OPEN PROBLEMS}
\bigskip
\bigskip
\centerline{Elliott H. Lieb\footnote{*}{Work partially
supported by U.S. National Science Foundation grant PHY~9019433 A02. To
appear in the proceedings of the conference ``Advances in Dynamical
Systems and Quantum Physics'', Capri, May, 1993 (World Scientific).
\copyright 1993 by the author.  Reproduction of this article, by any
means, is permitted for non-commercial purposes.}}
\centerline{{\it Departments of Mathematics and Physics, Princeton
University}}
\centerline{\it Princeton, NJ  08544  USA}
\bigskip
\bigskip
\centerline
{\sl Dedicated to Gian Fausto Dell'Antonio on his sixtieth birthday}
\bigskip
{\narrower\smallskip
\centerline{\bf Abstract}

The Hubbard model of interacting electrons, like the
Ising model of spin-spin interactions, is the simplest possible model
displaying many ``real world'' features, but it is much more difficult to
analyze qualitatively than the Ising model.  After a third of a century
of research, we are still not sure about many of its basic properties.  This
mini-review will explore what is known rigorously about the model and
it will attempt to describe some open problems that are possibly within
the range of rigorous mathematical analysis.\smallskip}
\bigskip
\bigskip\noindent
{\bf 0.  Introduction}
\medskip

The Hubbard model is to the problem of electron correlations as the Ising
model is to the problem of spin-spin interactions;  it is the simplest
possible model displaying many ``real world'' features.  It is, however,
much more
difficult to analyze qualitatively than the Ising model.  After a third of
a century of research, we are still not sure about many of its basic
features.  It is believed, for example, to have something to do with high
temperature superconductivity, and it would be nice to validate this.

One of the most intriguing questions concerns magnetism --- particularly the
magnetic properties of the ground state.  The familiar models of
interacting spins, such as the Ising or Heisenberg models, posit an
underlying spin Hamiltonian that ultimately comes from a model of
itinerant electrons.  The mystery here is that neither the electronic
kinetic nor the electronic potential energies favor ferromagnetism, but
together they sometimes do so.
Ferromagnetic spin models, for example, are popular, but the
truth is that the antiferromagnetic preference of the kinetic energy
usually seems to dominate.
In fact the only known examples of saturated itinerant electron
ferromagnetism come from the Hubbard model in a special limit (infinite
repulsion and one hole) devised by Nagaoka [NY] or from the Hubbard
model with finite repulsion but on special lattices devised by
Mielke and Tasaki [Mi1, TH2, MT] for which the kinetic energy spectrum is
macroscopically degenerate.  No doubt, further study of the
Hubbard model will eventually lead to a better understanding of the way in
which the Pauli exclusion principle leads to magnetism.

These notes will explore what is known rigorously about the model and it
will attempt to describe some open problems that are possibly within the
range of rigorous mathematical analysis.  The notes are not a complete
review in the sense that every contribution is covered, but an attempt has
been made to mention, at least, most topics that can be treated rigorously.
I apologize to those authors whose works have been omitted and I hope this
will be attributed to my ignorance rather than intent.  Given more space I
would have liked to discuss the closely related Falicov-Kimball model, for
which many rigorous results are known, but which is not a Hubbard model
because it does not have $SU(2)$ symmetry.  Another interesting chapter
would be the infinite dimensional Hubbard model---pioneered by Metzner and
Vollhardt [MV]---and which opens exciting mathematical and physical
avenues.  It has led to a large literature, but much remains to be
added in the way of mathematical rigor.

I thank Peter Eckle, Walter Metzner and Hal Tasaki for critically reading
this manuscript and I thank B\'alint T\'oth for considerable help with
the bibliography.
\bigskip
\bigskip\noindent
{\bf 1.  Definition of the Model}
\medskip

Only the original short-range Hubbard model will be considered
here.
Like the Ising model, the Hubbard model is defined on a graph, i.e., a
collection of vertices or sites (denoted by $\Lambda$ and whose number is
$\vert \Lambda \vert$) and (unordered) edges or bonds connecting
certain distinct pairs of vertices.
The word graph instead of lattice is
used to avoid any possible implication of translation invariance because
most of the results stated here do not depend on such invariance.
One is given a {\bf hopping
matrix} $T$, with elements $t_{xy}$, with $x$ and $y \in \Lambda$, and
we assume, as a convention, that  $t_{xy} = 0$ if $x,y$ are
not connected by an edge.  Note
that $t_{xx} = 0$. $t_{xy}$ might be complex, signaling the presence
of a magnetic field, the line integral of whose vector potential from
$x$ to $y$ (thought of now as points in $\R^3$) is $\arg (t_{xy})$.
However, $T$ is always self adjoint, $t_{xy} = t^*_{yx}$, with
$*$ denoting complex conjugate.

The {\bf bipartite} graphs form an important sub-class; here $\Lambda = A
\cup B$, with $A$ and $B$ disjoint, and such that there is no edge between
$x,y$ if $x \in A$ and $y \in A$ or if $x \in B$ and $y \in B$.  The square
lattice is bipartite, the triangular is not.

Electrons, i.e., spin $\mfr1/2$ fermions, move on $\Lambda$ with kinetic
energy given in second quantized form by $K = K_\uparrow + K_\downarrow$
with
$$K_\sigma = - \sum \limits_{x,y \in \Lambda} t_{xy} c^\dagger_{x
\sigma} c^{{\phantom{\dagger}}}_{y \sigma} . \eqno(1.1)$$
Here $\sigma
= \pm 1$ denote the two spin states $\uparrow$ and $\downarrow$ while
$c^\dagger_{x \sigma}$ is the creation operator for an electron at $x$
with spin $\sigma$.  We have $ c^\dagger_{x \sigma}
c^{{\phantom{\dagger}}}_{y \tau} + c^{{\phantom{\dagger}}}_{y
\tau}c^\dagger_{x \sigma}= \delta_{xy} \delta_{\sigma \tau}$ and $ c_{x
\sigma} c_{y \tau} + c_{y \tau}c_{x \sigma} = 0$. The number operator is
defined by $n_{x \sigma} = c^\dagger_{x \sigma}
c^{{\phantom{\dagger}}}_{x \sigma}$ and has eigenvalues 0 and 1.  The total
number of each spin species $$N_\sigma = \sum \limits_{x \in \Lambda}
n_{x \sigma}$$ is a conserved quantity.  The total particle number is $N =
N_\uparrow + N_\downarrow$ which satisfies $0 \leq N \leq 2 \vert \Lambda
\vert$.  The {\bf half-filled band}, $N = \vert \Lambda \vert$ is
especially important and especially amenable to analysis.

In the physics literature it is often assumed that $t_{xy} =$
constant $= t>0$ on all edges of $\Lambda$, in which case $T$ is a
discrete version of the Laplacian, but without the diagonal terms.
This assumption will not generally be made here.

At each site $x$ there is also given a number $U_x$ which governs the
on-site electron-electron interaction at $x$.  We usually assume all
$U_x \geq 0$ (repulsive case) or all $U_x \leq 0$ (attractive case).
The total potential energy is\footnote{$^\dagger$}{ The usual
formulation is $n_\uparrow n_\downarrow$ instead of our $(n_\uparrow -
\mfr1/2) (n_\downarrow - \mfr1/2)$.  If $U_x$ varies with $x$ the two
formulations are obviously inequivalent.  The formulation here is not
without a physical foundation because a neutral atom (with $n_\uparrow
+ n_\downarrow = 1$) is locally the most stable configuration and
adding or removing an electron produces a net local charge that raises
the energy roughly equally.} $$W = \sum \limits_{x \in \Lambda}
U_x (n_{x \uparrow} - \mfr1/2) (n_{x \downarrow} - \mfr1/2).
\eqno(1.2)$$ The total Hubbard Hamiltonian of our system is then $$H =
K + W, \eqno(1.3)$$ and most of the results discussed here are about
the ground state of this $H$.  In fact, many of the results do not
require a point interaction as in (1.2); instead, terms like $U_{xy}
(n_{x \uparrow} - \mfr1/2) (n_{y \downarrow} - \mfr1/2)$, with the
matrix $U_{xy}$ being positive semidefinite, are also allowed.

The interaction in (1.2) includes a one-body term $-\mfr1/2
\sum\nolimits_x U_x (n_{x \uparrow} + n_{x \downarrow})$ plus a trivial
constant term $\mfr1/4 \sum \nolimits_x U_x$.  If $U_x$ is independent
of $x$, as is normally assumed in almost all papers on the subject,
then this one-body term is trivially a constant proportional to the
fixed particle number.  It is interesting to consider $U_x \not=$
constant, and the one-body term is included in (1.2) in order to be
able to exploit {\bf hole-particle symmetry}:  The unitary
transformation that maps $c^{{\phantom{\dagger}}}_{x \downarrow}
\rightarrow c^\dagger_{x \downarrow}$ and $c^\dagger_{x\downarrow}
\rightarrow c^{{\phantom{\dagger}}}_{x \downarrow}$ (but $c_{x \uparrow}
\rightarrow c_{x \uparrow}$) is the hole-particle
transformation on the down-spins and it maps $(n_{x\downarrow} - \mfr1/2)$
into $-(n_{x\downarrow} - \mfr1/2)$.  Thus, this transformation maps the
repulsive $W$ into the attractive $W$ and vice versa, but it changes
the down-spin number from $N_\downarrow$ to $\vert \Lambda \vert -
N_\downarrow$.

This hole-particle transformation may or may not map $K_\downarrow$ into
$K_\downarrow$.
It does map $K_\downarrow$ into $\widehat K_\downarrow$, defined by $\widehat
t_{xy} = - t^*_{yx}$.

In the special, but important case that $T$ is real and $\Lambda$ is
bipartite we can make a further unitary transformation that will take
$\widehat K_\downarrow$ into $K_\downarrow$ and $n_{x \downarrow}$ into
$n_{x \downarrow}$.  This unitary transformation
maps $c_{x \downarrow}$ into $(-1)^x c_{x
\downarrow}$ (and, of course, $c_{x \uparrow}$ into $c_{x \uparrow}$),
where $(-1)^x$ denotes the function on the vertices of $\Lambda$ which is
$+1$ for $x \in A$ and $-1$ for $x \in B$.  Effectively,
$t_{xy}$ is mapped into $(-1)^x (-1)^y t_{xy} = -
t_{xy}$ for the down-spins.  Henceforth, the hole-particle transformation
on bipartite lattice is always meant to include this additional unitary
$(-1)^x$, so that $K_\downarrow$ is mapped into $K_\downarrow$.

Thus, in the special real, bipartite case we can analyze the repulsive case
by analyzing the attractive case.  But we have to remember that
$N_\downarrow \leftrightarrow \vert \Lambda \vert - N_\downarrow$ so that
the particle number for one is related to the magnetization of the other.
This is a non-trivial distinction and is similar to the relation between
the ferromagnetic and antiferromagnetic Ising models on a bipartite graph.

The hole-particle transformation also induces a conservation law in the
real, bipartite case that has no classical analogue.  Like any electron
system this model has an $SU(2)$, i.e., angular momentum, invariance.  The
generators are
$$J^3 = \mfr1/2 (N_\uparrow - N_\downarrow), \qquad J^+ = \sum \limits_{x \in
\Lambda} c^\dagger_{x\uparrow} c^{{\phantom{\dagger}}}_{x \downarrow},
\qquad J^- = (J^+)^\dagger. \eqno(1.4)$$
However, the Hamiltonian $H$ in (1.3) is the unitary transform (described
above) of another
$H^\prime$ (with $U_x$ replaced by $-U_x$) which also has an $SU(2)$
symmetry with generators given by (1.4) in the {\it transformed} basis.  By
transforming back we obtain a different set of $SU(2)$ generators
$$\widehat J^3 = \mfr1/2 (N_\uparrow + N_\downarrow - \vert \Lambda \vert),
\qquad \widehat J^+ = \sum \limits_{x \in \Lambda} (-1)^x c^\dagger_{x
\uparrow}
c^\dagger_{x \downarrow}, \qquad \widehat J^- = (\widehat J^+)^\dagger
\eqno(1.5)$$
that also commute with $H$ and with the $J$'s in (1.4).  We call this the
{\bf pseudospin}.  This special model thus has an $SU(2) \times SU(2)$
symmetry group,\footnote{$^{\S}$}{ Actually, the group is
$SO(4) = SU(2) \times SU(2) /Z_2$, as pointed out by Yang and Zhang
[YZ].  The reason is that the two operators $w = -{\bf 1}$ in each
$SU(2)$ corresponds to only {\it one} operator on our Hilbert space,
i.e., $w \otimes \1 = \1 \otimes w$.  This reduction to $SO(4)$
coincides with the observation that in every state the spin and
pseudospin are either both integral or both half-integral.}
but it must be emphasized that real $T$ is {\it essential} (zero magnetic
field).  These operators change the quantum numbers $N_\uparrow$ and
$N_\downarrow$.  The operators $J^\pm$ change $J^3$ by one unit while
$\widehat J^\pm$ change $\widehat J^3$ by one unit.  The four operators
together thus permit us to move around inside rectangles in
$(N_\uparrow, N_\downarrow)$ space whose four vertices have the form
$(n,m), (m,n), (\vert \Lambda \vert - n, \vert \Lambda \vert - m),
(\vert \Lambda \vert - m, \vert \Lambda \vert - n)$.  This means that
to each eigenstate for $N_\uparrow = n$, $N_\downarrow =m$ there is a
corresponding eigenstate with the same energy at each point in the
rectangle (there are also additional eigenstates, of course).  Thus,
the usual operators $J^\pm$ permit us to infer all the states of our
Hamiltonian from knowledge of the states on the line $N_\uparrow -
N_\downarrow =0$ or the line $N_\uparrow - N_\downarrow = 1$.  The
pseudospin operators $\widehat J^\pm$ permit us to infer everything
{}from knowledge of the line $N_\uparrow + N_\downarrow = |\Lambda |$
(half-filled band) or the line $N_\uparrow + N_\downarrow = \vert
\Lambda \vert + 1$.  But we repeat that this property of $\widehat
J^\pm$ holds {\it only} for the real, bipartite case.

We note, for future use, that the hole-particle transformation ({\it
without} $(-1)^x$) applied to {\it both} spins (namely
$c^{{\phantom{\dagger}}}_{x \sigma}
\leftrightarrow c^\dagger_{x \sigma}$).  does not preserve $\vec J$.
Indeed, $J^3 \rightarrow - J^3, J^\pm \rightarrow - J^\mp$, but $\J^2
\rightarrow \J^2$.

The Hubbard model describes --- in the simplest possible fashion --- an
interacting fermion system.  It can be viewed this way, as a toy model,
or it can be viewed, a bit more realistically in the repulsive case, as
a serious model of $\pi$-electrons hopping between localized Wannier
orbitals in some molecule such as benzene (with $\vert \Lambda \vert =
6$); the half-filled band, $N = \vert \Lambda \vert$, is then
especially important because it corresponds to neutrality.  The
ultra-short range interaction is supposed to mimic a highly screened
Coulomb potential.  From the latter viewpoint it was known first in the
chemistry literature as the Pariser-Parr [PP]-Pople [PJ] model;
molecules having a bipartite structure are called ``alternant
molecules''.  It was a decade later that Hubbard [HJ], Gutzwiller [GMC]
and Kanamori [KJ] realized its importance for bulk matter.
\bigskip
\bigskip\noindent
\noindent
{\bf 2.  One-Dimensional Exactly Solvable Model}
\medskip

In 1968 the ground state of $H$ was solved for the translation invariant
one-dimensional ring by Lieb and Wu [LW] using the extension of the
``Bethe ansatz'' technique [BH, LL1] to fermions [MJ, FL1, YC, GM].
Shortly thereafter, Ovchinnikov [OA] used these results to calculate the
elementary excitation spectrum at half filling.  Recently, Essler and
Korepin [ES] obtained a new and illuminating derivation.  Coll [CC]
extended Ovchinnikov's results
to arbitrary filling, as did Woynarovich [WF].  Takahashi [TM]
evaluated the magnetism curve at half filling and Shiba [SH] evaluated
the magnetic susceptibility for all filling.  Koma [KO] has formulated a
sequence of approximations to the positive temperature free energy and
correlation length (obtained by using the Trotter product formula for the
partition function) which converge to the exact answer and such that each
approximation can be calculated using the Bethe ansatz without any
assumptions, such as the ``string hypothesis''.

These results were obtained in the thermodynamic limit, in which sums
could be replaced by integrals.  Woynarovich and Eckle [WE] evaluated the
asymptotics of finite size effects on the ground state energy.
For small chains the general ``Bethe
ansatz" solution, while correct, is too complicated for numerical
evaluation.  Heilmann and Lieb
[HL]  undertook to evaluate {\it all} the energy levels for {\it all}
$U > 0$ for the benzene molecule ($N = 6, \vert \Lambda \vert =6$).  To
our surprise we found many instances of both level crossings and of
permanent degeneracy --- as a function of $U$ --- and which were not
accounted for on the basis of the known invariance groups (spin,
pseudospin and symmetries of the hexagon).  This means that the system
has {\it non-abelian} symmetry groups and these {\it are dependent on}
$U$ (i.e., the group operations commute with $H$ but {\it not} with $K$
and $W$ separately; a well known example of this phenomenon, for the
hydrogen atom, is the Runge-Lenz vector whose definition depends on the
value of the electron's charge).

What are these invariants?  Sixteen years later Shastry [SB] (unaware
of [HL]) found many invariants and also a two-dimensional classical
statistical mechanics vertex model whose transfer matrix commutes with
our $H$.  In fact he found a whole commuting family of such transfer
matrices --- which means that the one-dimensional model can be called
``integrable''.  It is not clear whether all invariants of $H$ are of
Shastry's form.  A few years later, Grosse [GH] (motivated by [HL])
published another derivation of some of the invariants.

Another interesting question is whether all the eigenstates of $H$ are
of the ``Bethe ansatz'' form.  This was answered in the negative by
Essler, Korepin and Schoutens [EKS] who went on to demonstrate that
when the ring is bipartite (i.e., even length) the $SU(2) \times SU(2)$
generators $J^+$ and $\widehat J^+$ save the day.  The lowest weight
states (the ones that are annihilated by $J^-$ and $\widehat J^-$) are
claimed to be {\it all} of the ``Bethe ansatz'' type.  All the
remaining states are then obtained by application of $J^+$, and
$\widehat J^+$.

Many of the ``Bethe Ansatz'' results rely on a ``string'' hypothesis.
Moreover, some assertions in [EKS], notably the linear independence of
the solutions, rely on an appeal to some properties of Shastry's
invariants that have not been verified. It is desirable to put these
matters on a more rigorous basis.

The literature about this one-dimensional solution is vast and the
above remarks do not reflect everything that is known about the
subject.

\bigskip
\bigskip\noindent
{\bf 3.  Magnetism}
\medskip\noindent
{\bf A.  One-Dimension.}
\smallskip

It is convenient, now, to take an open chain
instead of a ring.  Then, by a general theorem of Lieb and Mattis [LM1],
whose proof also works for lattice systems with nearest neighbors hopping
and a {\it completely arbitrary} many-body potential, $E^N (S)$, the ground
state energy of $H$, as a function of total spin $S \leq \mfr1/2 N$ and for
$N$ particles, satisfies
$$E^N (\mfr1/2 N) > E^N (\mfr1/2 N - 1) > E^N (\mfr1/2 N - 2) > \dots >
E^N(0) \ \hbox{or} \ E^N(\mfr1/2). \eqno(3.1)$$
The numbers $U_x$ are now totally arbitrary and not necessarily of one
sign.  [Note:  The original theorem [LM1] uses a Perron-Frobenius
positivity argument, which a-priori requires $t_{xy} > 0$. However, any
$T$ can be reduced to this form in one-dimension
by the simple unitary gauge transformation
$c_{x\sigma}\rightarrow \exp[-i\theta(x)]c_{x\sigma}$,
which carries $t_{xy}$ into $\widehat t_{xy} :=
t_{xy} \exp [i \theta (x) - i \theta (y)]$, where $\theta (x)$ is an
arbitrary real number, selected so that $\widehat t_{xy} > 0$ for all
$x$ and $y$.]
Since we can then take $T$ to be real, our $H$ is then the (down-spin)
hole-particle transform of some other $\widehat H$, in which $U_x$ is replaced
by $-U_x$.  Assuming $N$ to be even, (3.1) says that the ground state of
$H$ has $N_\uparrow = N_\downarrow = N/2$.  The transformed
$\widehat H$ system has $\widehat N_\uparrow = N_\uparrow,\ \
\widehat N_\downarrow = \vert \Lambda \vert - N_\downarrow$, whence
$\widehat S^3 = \mfr1/2 (\widehat N_\uparrow - \widehat N_\downarrow)$ is the
predetermined number $\mfr1/2 (N - \vert \Lambda \vert)$.  The $\widehat H$
system also satisfies (3.1) since it is one-dimensional, and the spin of
the ground state of the $\widehat H$ system is therefore
$$\widehat S = \mfr1/2 \big\vert N- \vert \Lambda \vert \big\vert.$$
This number, $\widehat S,$ is
{\it the pseudospin of our ground state} --- a curious result whose physical
significance is not entirely obvious.  (Note the logic here.  Both systems,
$H$ and $\widehat H$ must be in their respective ground states, consistent
with the given conditions on each; for $H$ it is $N_\uparrow +
N_\downarrow = N$ and for $\widehat H$ it is $N_\uparrow - N_\downarrow = N
- \vert \Lambda \vert$.)

This theorem was extended many years later by Aizenman and Lieb [AL] to
positive temperatures.  The main theorem, applicable to any many-body
potential in a one-dimensional system, expresses the fact that the free
energy is a monotone increasing function of the spin.  This is done in terms
of total spin $S$ or $3$-component $s$, as follows; neither inequality
implies the other.

Classify the eigenstates by the 3-component of spin, $J^3 = s = \mfr1/2
(N_\uparrow - N_\downarrow)$, and by $S$, the total spin angular momentum
(recall $\J^2 = S(S+1)$).  For a given $N$ and inverse temperature $\beta$,
let $Z_3 (s) = \Tr_s e^{-\beta H}$ be the partition function in which only
states of a given $s$ value are included.  Likewise, let $Z_J (S) =
\Tr_S e^{-\beta H}$ be the partition function for a given $S$.  We
have the relation $Z_3 (S) - Z_3 (S+1) = (2S+1)^{-1} Z_J (S)$, which is
obvious from the theory of angular momentum.  Correspondingly, we have the
combinatorial quantities $Y_3 (s) := {N \choose {N \over 2} + s}$ and
$Y_J (S) := (2S+1) [Y_3 (S) - Y_3 (S+1)]$, which are essentially the
partition functions of free particles with $H = 0$, and which serve to
normalize the $Z$'s.  The two theorems are then that
$${Z_3 (s) \over Y_3 (s)} \ \hbox{\quad\sl and \quad} \
{Z_J (S) \over Y_J (S)}$$
are {\it both strictly monotone decreasing functions of their arguments}
($s$ or
$S$).  A corollary of this is that the magnetization is less than
its value in the {\bf atomic limit} (or what I prefer to call the pure
paramagnetic value).  I.e., for all $\beta$ and magnetic field $h$
$$M(\beta, h) = {1 \over \beta} {\d \over \d h} \ln Z (h) < N \tanh (\beta
h) . \eqno(3.2)$$
Here $Z(h) = \sum \limits^{N/2}_{s=-N/2} Z_3 (s) \exp [2 \beta sh]$ is
the total partition function (recall that the $g$-factor of an electron is
2).

It is noteworthy that all this holds for {\it completely arbitrary},
Hermitian $t_{xy}$ and (real) $U_x$.  Indeed, it holds even if we add an
arbitrary real one body potential
$$V = \sum \limits_{x \in \Lambda} V_x (n_{x\uparrow} + n_{x \downarrow}).
$$

An amusing fact about one-dimension concerns the $U = +\infty$ case.  For an
open chain the ground state is highly degenerate---indeed, it can have any
value of $S$.  For a closed chain the situation is quite different and the
$S$ of the ground state depends on the sign of the $t_{xy}$'s and whether
$N$ is even or odd; $S$ can be $N/2$ in some cases [AL].  The situation is
discussed in detail by Mielke [Mi2] who finds that the ``average" $S$ is
$\sqrt{N}$.  Some interesting facts about the closed one-dimensional chain
with $U = 0$, but with a magnetic field, can be found in [LL2].  The $U
\not= 0$ case is discussed by Fujimoto and Kawakami [FK1].
\medskip\noindent
{\bf B.  Half-Filled Band.}
\smallskip

The hole-particle symmetry notwithstanding,
the repulsive case (all $U_x \geq 0$) and the attractive case (all $U_x
\leq 0$) are quite different, even for a bipartite lattice.
The physical spin of one is the pseudospin of the other.

In the limit $U \rightarrow + \infty$ (by which I mean all $U_x \rightarrow
+ \infty$), the energy and wave functions have nice limits.  The electrons
become hard-core particles.  When $N = \vert \Lambda \vert$ we just have
one electron per site and, since motion is impossible, each electron can
be, independently, spin-up or spin-down.  This gives us
the atomic limit
whose partition function is $Y_3$ or $Y_J$ given above.
First order perturbation theory in $t/U$ vanishes, but in second order we
have to diagonalize our $H$ among all the $2^{\vert \Lambda \vert}$
degenerate ground states just described.  This yields an effective
Hamiltonian $H^\prime$ that can be written in terms of the three Pauli spin
operators $\bs$ at each site.  It turns out [AP] that for any graph
$$H^\prime = \sum \limits_{x,y \in \Lambda} J_{xy} (\bs_x \cdot \bs_y -
\mfr1/4) \eqno(3.3)$$
with $J_{xy} = \vert t_{xy} \vert^2 (U^{-1}_x + U^{-1}_y)$.  This is the
spin 1/2
antiferromagnetic Heisenberg Hamiltonian and it is known [LE1, LM2] that its
ground state has total spin
$$S = \big\vert \vert A \vert - \vert B \vert \big\vert \eqno(3.4)$$
on a bipartite graph.  It is also known [DLS, KLS] that in the {\bf
translation invariant case} (i.e., our graph is a $D$-dimensional hypercube
with periodic boundary conditions and $t_{xy} =$ constant $=t$ and $U_{xy}
=$ constant $=U$) there is long range order when $D \geq 3$.

The obvious question is whether the results just stated (i.e., (3.4) and
the long range order) hold non-perturbatively in the repulsive case.  One
would also guess that in the attractive case the total spin should be zero
in the ground state because when $U_x = - \infty$ for all $x$ the ground
state consists simply of bound pairs of electrons sitting on selected
sites.  These questions about the spin are answered in the
following [LE1].
\smallskip

{\bf Theorem 1:}  {\it Assume $t_{xy}$ is real for all $x,y \in
\Lambda$.  If $U_x < 0$ for all $x$, the ground state on any connected
graph is unique and has spin $S = 0$} for any even electron number, $N$, not
just $N = \vert \Lambda \vert$.  {\it If $U_x
> 0$ for all $x$, if $\Lambda$ is connected and bipartite and if $N = \vert
\Lambda \vert$ is even, the ground
state is unique (except for the $(2S +1)$-fold degeneracy) and has spin $S
= \mfr1/2 \big\vert \vert A \vert - \vert B \vert \big\vert$.}
\smallskip

We can easily have $\vert A \vert - \vert B \vert$ of the order of $\vert
\Lambda \vert$ itself.  As an example, take a square lattice and add a site
at the center of each bond of this square lattice.  The original sites are
then $B$ sites and the new sites are $A$ sites.  Then $\vert A \vert = 2
\vert B \vert$ and the ground state has a bulk magnetization per site of
1/3.  This is really more like ferrimagnetism than ferromagnetism but, in
any case, it is one of the few examples known in which the system has a
bulk magnetization without an external magnetic field.

There is an interesting corollary of this theorem if $\Lambda$ is bipartite
and if $\vert A \vert \geq \vert B \vert$.  Suppose we ask for the absolute
minimum energy, without fixing $N$.  Starting with $U_x > 0$ we find that
an optimum $N$ is $N = \vert \Lambda \vert$ (by the $SU(2) \times SU(2)$
symmetry mentioned in Sec.~1).  But $J^3 = (N_\uparrow - N_\downarrow)/2$
can be anywhere in the interval $(\vert B \vert - \vert A \vert)/2$ to
$(\vert A \vert - \vert B \vert) /2$ since $S = (\vert A \vert - \vert B
\vert)/2$.  Now using hole-particle symmetry to
obtain the $U_x < 0$ model, we find that the optimum $N$ is any integer
satisfying $2 \vert B \vert \leq N \leq 2 \vert A \vert$.  (E.g., starting
with $N_\uparrow = \vert A \vert, N_\downarrow = \vert B \vert$, the
transformed values are $N_\uparrow = \vert A \vert, N_\downarrow = \vert
\Lambda \vert - \vert B \vert$, which yields $N = 2 \vert A \vert$.)
Thus, {\it there can
be a large degeneracy in the attractive case!}  (For $U = 0$ this is easily
seen from the remark that $T$ has at least $\vert A \vert - \vert B \vert$
zero eigenvalues.)

This theorem was extended to positive temperature by Kubo and
Kishi [KK] who found upper bounds on certain two-point functions.  They
discuss only the translation invariant case on a hypercubic lattice
with $U_x = U =$ constant, but
their method easily extends to the general case.  For $U < 0$ they bound
the spin susceptibility at wave vector $q$ by
$$\upchi_q \leq {1 \over 4 \vert U \vert} \eqno(3.5)$$
for all temperatures and all filling fraction $N/\vert \Lambda \vert$ (more
precisely, they use the grand canonical ensemble and prove (3.5) for all
chemical potentials $\mu$).  This result precludes magnetic long range
order.  In the repulsive case $U > 0$, and with $\mu$ adjusted to the
half-filled
band case $N = \vert \Lambda \vert$, namely $\mu = 0$, they bound the
charge susceptibility as
$$(\delta \widehat n_q, \delta \widehat n_{-q}) \leq (\beta U)^{-1}
\eqno(3.6)$$
and the pairing susceptibility as
$$(\widehat p_q, \widehat p_{-q}) \leq (\beta U)^{-1}. \eqno(3.7)$$
Here $(A,B)$ is the Duhamel two-point function $(A,B) = \int^1_0 \Tr
[A^\dagger e^{t \beta H} B e^{(1-t)\beta H}]\ dt$ and $\delta \widehat n_q =
\widehat n_q - \langle \widehat n_q \rangle$ and $p_x = c^\dagger_{x
\uparrow} c_{x \downarrow}$ and $\widehat{\phantom{a}}$ denotes spatial Fourier
transform. Charge long range order is precluded by (3.6) while Cooper
pairing is precluded by (3.7).

Although I am restricting this review to the Hubbard model, I cannot resist
the temptation to mention that Theorem 1 has recently been extended
[FL2] to another model---the Holstein model---in which electrons interact with
a quantized phonon field instead of with each other.  Again, the finite system
ground state for any even number of electrons is unique and has zero spin.
The method of proof of Theorem 1 has also been used by Ueda, Tsunetsugu and
Sigrist [UTS] to show that the periodic Anderson model at half-filling has
a singlet ground state.

This close connection between the half-filled band, repulsive Hubbard
model and the antiferromagnetic Heisenberg model points to the first of
our spin problems.  The antiferromagnetic Heisenberg model on a
hypercubic lattice, in the thermodynamic limit, has no long range order
(LRO) at positive temperature in dimensions $D = 1$ or 2.  This is a
consequence of the Hohenberg-Mermin-Wagner theorem.  The same is true
for the Hubbard model as first shown by Walker and Ruijgrok [WR], then
by Ghosh [GD].  Later, Koma and Tasaki [KT]proved it by a different
method---that of McBryan and Spencer [MS].  As far as the ground state
is concerned, the Heisenberg model has LRO for $D = 2$ and spin 1 or
more per site [KLS] (i.e. $\vert \bs_x \vert^2 = S(S +1)$ and $S \geq
1$) and it is believed to have LRO also when $S = 1/2$, which is the
case of interest for us.  The $D =1$ case is believed to have no LRO in
the ground state.  For $D \geq 3$ there is LRO in the ground state for
all $S \geq 1/2$ [KLS] and at positive temperature [DLS, KLS] for $S
\geq 1$ (presumably, also for all $S \geq 1/2$).  The obvious
conjecture is the following.
\smallskip

{\bf Problem 1:}  {\it Prove that there is antiferromagnetic LRO in the
half-filled band,
repulsive Hubbard model on the hypercubic lattice (with $t_{xy} = 1$ for
$\vert x-y \vert = 1$ and $U_x =$ positive, finite  constant) in the
ground state
when $D = 2$ and for positive temperature when $D \geq 3$.  For which values
of $N$ will the attractive model have LRO?}
\smallskip

The Falicov-Kimball [FK2]
model poses an analogous problem that {\it can be solved} affirmatively.
In this model, one kinetic energy term, $K_\downarrow$, is omitted from
the Hamiltonian.  Although the down spins are not dynamic their locations
are left arbitrary.  After ``integrating out'' the movable particles
(up-spins), an effective interaction among the fixed particles (down-spins)
is left.  This, then, is a classical lattice gas with a complicated
interaction.  It
resembles an Ising (not Heisenberg) model when $U$ is large and it
can be shown [KL] to have long range order at low temperatures (and no long
range order at high temperatures) in two or more dimensions---as does the
Ising model.
\medskip\noindent
{\bf C. The Surprising Hole.}
\smallskip

In the previous subsection we considered the
half-filled band and showed a strong tendency to antiferromagnetism.  The
only important requirement on $T$ was that it was real (i.e., no magnetic
field acts on the orbital motion).  Nagaoka [NY] made a surprising
discovery about the case $U = + \infty$, but {\it with one hole\/} (i.e.,
$N = \vert \Lambda \vert - 1$).  Thouless [TD] had a similar result a bit
earlier (cf. note 7 in [NY] and the discussion on p. 47 in [LE2]) but there
is little doubt that Nagaoka's presentation of this particular result is
clearer, more precise and applicable to certain non-bipartite lattices such
as $bcc$ and $fcc$.  The Thouless approach uses a Perron-Frobenius argument
that appears to be restricted to bipartite lattices; for such lattices it
is true that fermions behave like bosons when $U = + \infty$ and there is
one hole.  Nagaoka, on the other hand seems to require a regular Bravais
lattice, but this is not really necessary provided {\it all\/} $t_{xy}$ are
nonpositive.  The fully general result with a considerably simplified
proof, was given by Tasaki [TH1].

It is usually assumed in the condensed matter literature that the
$t_{xy}$'s are nonnegative, but there seems to be neither a compelling
reason for this assumption nor many examples in which it can be verified
(S. Kivelson, private communication).  (For a bipartite lattice, however,
one can have either sign with the help of the unitary operator $(-1)^x$
applied to {\it both} spin $\uparrow$ and $\downarrow$.)
Notice that $t_{xy} > 0$ puts the lowest kinetic energy at one point,
namely $k=0$, in Fourier space.  The condition $t_{xy} < 0$ puts it at the
edge of the Brillouin zone, and thus the lowest kinetic energy can be
highly degenerate.  The physical intuition is then quite different in the
two cases---a fact that should not be lost sight of because of the
hole-particle symmetry that holds for bipartite graphs.  The proof in [TH1]
shows that negative, not positive
$t_{xy}$ is the {\it natural mathematical assumption\/} for this theorem.
If this upsets anyone's physical proclivity, that is a pity.
\smallskip

{\bf Theorem 2:} {\it If $N = | \Lambda | - 1$, if $U_x = +\infty$ for all
$x\epsilon \Lambda$, and if $t_{xy} \le 0$ for all $x,y$,
then the ground state has total spin $S = N/2$.  This state is unique
up to the trivial $(N+1)$--fold degeneracy if $\Lambda$ satisfies a
certain connectivity condition [TH1].}
\smallskip

The connectivity condition mentioned above is not stringent and it holds
for all regular lattices in dimension greater than one (see also [AL]).
Essentially it means that there are loops that permit nontrivial
permutations of the particles.

In the case that $\Lambda$ is completely translation invariant, i.e., that
all vertices of $\Lambda$ are equivalent, as is the case on a hypercubic
lattice on a torus, a different proof of the theorem, very similar to
Nagaoka's, was given by Tian [TG1] and by Trugman [TS].

The obvious next question to ask is this: {\it If there is more than one
hole $(N < |\Lambda| - 1)$ and $U = +\infty$, is the ground state totally
ferromagnetic, i.e., is $S = N/2$?}  There can be no simple general theorem
because numerical calculation on small systems show that the answer seems
to be ``no" and, at the same time, no simple pattern seems to emerge.  Yet
there are a few theoretical results, as follows.

(i) The Nagaoka-Tian-Trugman method can be generalized for $\Lambda =$ the
\hfill\break $D$-dimensional
hypercubic lattice with periodic boundary conditions and with $t_{xy} = t
=$ nonpositive constant to show [TG2, TG3, TS, SQT] that the completely
magnetized state
energy,
$E (S = N/2; \Lambda)$, when compared to the actual ground state
energy, $E(\Lambda)$, satisfies
$$
\lim_{\Lambda \to \infty} E (S = N/2; \Lambda) - E (\Lambda) = 0
\eqno(3.8)$$
in the thermodynamic limit, $\Lambda \to \infty$, when the number of
holes $N_h = |\Lambda| - N$ is not too large.  The best result is by
Shen, Qiu and Tian [SQT], which gives (3.8) when $N_h <
|\Lambda|^\alpha$ with $0 < \alpha < 2 / (D+2)$.  Note that we do not
divide by $|\Lambda|$ in (3.8), which thus truly represents the
vanishing of a gap.  The proof here is elegant and simple. However, one
would expect (3.8) to hold as long as $N_h / |\Lambda| \rightarrow 0$
as $|\Lambda| \rightarrow 0$.

(ii) If there are many holes, $N_h / |\Lambda| > 0.49$ for the $D=2$ or
$N_h / |\Lambda| > 0.32$ for the $D=3$ hypercubic lattices, and $\Lambda
\to \infty$ as in (i) then
$$
E (S = N/2; \Lambda) - E (S = N/2 - 1; \Lambda) \not\rightarrow 0.
\eqno(3.9)$$
There really is an instability of the $S = N/2$ state with respect to
one spin flip.  This was proved by Shastry, Krishnamurthy and Anderson
[SKA]; see also [SA1]. The estimate was improved to $N_h / |\Lambda| >
0.29$ by von der Linden and Edwards [LvE]; Hanisch and
M\"uller-Hartmann [HM] simplified the calculation (but not the estimate
of 0.29).

(iii) Several authors [DW, FRDS, SA1, TB] were able to prove, for a
translation invariant hypercubic lattice model, that when there are two
holes the ground state energies satisfy $E (S = N/2 - 1; \Lambda) < E (S = N/2;
\Lambda)$.  However, assumptions have to be made about the relative lengths
of the sides.  S\"ut\H o [SA1] extended this to $2, \dots , 6$ holes for a
bcc lattice.

(iv)  S\"ut\H o [SA2] shows, as expected, that the energy splitting needed for
demagnetization is, in any case, negligible.  He proves that $M(\beta, h,
\uprho)$, the magnetization per site in field $h$ at density $\uprho =
N/\vert \Lambda \vert$ satisfies (in the thermodynamic limit) $M(\beta, h,
\uprho) \rightarrow \tanh (\beta h)$ as $\uprho \rightarrow 1$.

The results in (ii) and (iii) are achieved with a variational calculation.
The value of $E (S = N/2; \Lambda)$ is easy to find exactly because it
equals the energy of spinless electrons, i.e.,
$$
E (S = N/2; \Lambda) = \sum^N_{j=1} \lambda_j (T) \eqno(3.10)
$$
for any $\Lambda$ and hopping matrix, T, and in which $\lambda_1 (T) \leq
\lambda_2 (T) \leq \cdots $ are the eigenvalues of T.  Thus, the hard problem
is to find
a good variational function with $S = N/2 - 1$, and this appears to be
extraordinarily difficult.  Why? No one seems to know!  And why is it so
difficult to treat $S = N/2 - 2$?

These results, (i)--(iii), show that one cannot expect $S = N/2$
except when $N = \vert \Lambda \vert - 1$, but one can ask the following.
\smallskip

{\bf Problem 2:}  {\it With $U = + \infty$, for which $\uprho := N/\vert
\Lambda \vert$ is it true that some (if there is more than one) ground
state has $S/\vert \Lambda \vert > 0$ in the thermodynamic limit $\Lambda
\rightarrow \infty$?}
\smallskip

This brings us to two more open problems about the $U = +\infty$ case; the
first
is a corollary of the second.
We take a large $\Lambda$ and $N$ particles and suppose that the thermodynamic
limit $\Lambda \to \infty$ with $\rho = N/|\Lambda|$ fixed is well defined.
We set $S =$ spin of the ground state (the maximum such spin if there is
more than one ground state) and we set $S_{\max} = N/2$.  We also assume
that $t_{xy} \geq 0$ for all $x$ and $y$.  Then
\smallskip

{\bf Problem 3}: {\it Prove or disprove that}
$$
\lim_{\rho \to 0} \ \lim_{\Lambda \to \infty} \ \ S / S_{\max} = 0.
\eqno(3.11)$$

\smallskip
{\bf Problem 4}: {\it Does there exist some number $\rho_c > 0$ such that}
$$
\lim_{\Lambda \rightarrow \infty} \ S / S_{\max} = 0 \ \ {\sl for \ all} \
\rho < \rho_c \, ?\eqno(3.12)$$
\smallskip

The requirement that $t_{xy} \geq 0$ is important.  As we shall see in
Sect.~D, Mielke's work shows that there can be nice, periodic lattices in
any dimension (such as the kagome lattice in two-dimensions) for which $S =
S_{\max}$ for all $\uprho < \uprho_c$ with $\uprho_c > 0$, thereby
contradicting (3.11) and (3.12).  To achieve this, however,
one needs $t_{xy} \leq 0$.  Perhaps (3.11) and (3.12) hold in the case
$t_{xy} \leq 0$ if we replace ``maximum such spin'' by ``average of such
spins'', because Mielke's and Tasaki's examples have highly degenerate
ground states with spins ranging from 0 to $N/2$.

Closely related in spirit to Theorem 2, but with
an interesting, different proof, is the result of Chakravarty, Chayes and
Kivelson [CCK]. They start with a half-filled band, $N = |\Lambda|$, and $U$
large.  Then they add or subtract one or two particles, and define
$E_n: = E (N = |\Lambda| + n)$ for $-2 \leq n \leq +2$.  They then define
$$
\Delta_e = 2E_1 - E_2 - E_0$$
$$
\Delta_h = 2E_{-1} - E_{-2} - E_0 . \eqno(3.13)$$
The interpretation of $\Delta_e$ is as a 2-particle binding energy, while
that of $\Delta_h$ is as a 2-hole binding energy.  The picture of
$\Delta_e$, for example, is that given two very large systems at
half-filling, and given two extra electrons, is it energetically favorable
to add the two electrons to one system $(\Delta_e > 0)$, or is it favorable
to add one electron to each system $(\Delta_e < 0)$?  The former, $\Delta_e
> 0$, connotes {\bf pair binding}.

It is pointed out in [CCK] that $\Delta_e \leq 0$ and $\Delta_h \leq 0$
when $U = 0$, but they quote numerical studies showing that $\Delta_e >
0$ and $\Delta_h > 0$ for some $U$ and some $\Lambda$.  They prove,
however, that in the limit $U \rightarrow \infty$, $\Delta_e \leq 0$ if
all $t_{xy}$ are nonnegative and $\Delta_h \leq 0$ if all $t_{xy}$ are
nonpositive.  For a bipartite graph the sign does not matter (as long
as all $t_{xy}$ are positive or all are negative), and thus $\Delta_e
\leq 0$ and $\Delta_h \leq 0$ in this case.  They interpret this result
to mean that the numerical positive binding results are only an
intermediate $U$ phenomenon but, strictly speaking, the (unlikely) alternatives
$\Delta_e = 0$ and $\Delta_h = 0$ in their theorems would first have to
be eliminated.
\medskip\noindent
{\bf D.  Another Path to Ferromagnetism}.
\smallskip

The one-hole, $U = + \infty$ model is not the only one known to have {\it
saturated} ferromagnetism (i.e., $S = N/2$).  Mielke [Mi1] and later Tasaki
[TH2] and then both [MT] found interesting, but very special models with
this property.

The basic idea is to find a graph $\Lambda$ and a hopping matrix $T$ such
that the lowest eigenvalue of $T$ (call it $\lambda_0$) is highly
degenerate; in fact we want the degeneracy to be at least $N$ and, to be
interesting, we want that to be of the order of $\vert \Lambda \vert$.  Let
$N_0$ denote the degeneracy of this lowest eigenvalue and denote the space
of these eigenfunctions by $\ch_0$.  If $U_x \geq 0$ for all $x \in
\Lambda$, it is easy to see that if $N =
N_0$ then
\item{(i)} There is a ground state having $S = N_0/2$.
\item{(ii)} The state is simply a determinant formed from the $N_0$ vectors in
$\ch_0$ and its energy is $N_0 \lambda_0$.
\item{(iii)} The ground state is unique if certain additional conditions
(known to be optimal) are met [Mi1].

If $N < N_0$ the ground state manifold will contain at least one state with
$S = N/2$, but perhaps others as well (see [Mi1]).

A comparison with Theorem 1 is interesting, but it is not clear whether or
not it is misleading.  Note that the $S = \big\vert \vert A \vert - \vert B
\vert \big\vert /2$ result there for a bipartite graph and a half-filled
band was somehow related to the $\big\vert \vert A \vert - \vert B \vert
\big\vert$-fold degeneracy of the zero-mode of $T$.  Thus, a common feature
is degeneracy, and it is often said that itinerant ferromagnetism is
associated with atomic or kinetic energy degeneracy.  But there are also
important differences:
\item{(i)} The spin in Theorem 1, while it might be
proportional to $\vert \Lambda \vert$, is not $N/2 = \vert \Lambda \vert
/2$.
\item{(ii)} No fine tuning of $T$ or of $\Lambda$ was needed for Theorem 1.
All that was needed was the bipartite structure, the positivity of
the $U_{x}$'s and the reality of the $t_{xy}$'s.

Mielke's way of achieving the degeneracy $N_0$ is to start with some graph
$G$ and then to set $\Lambda = L(G)$, the {\bf line graph of G}, which is
defined as follows.  Make a mark in the center of every edge of $G$;  those
marks will be the sites of $L(G)$.  Two sites are connected by an edge in
$L(G)$ if the two edges of $G$ on which they reside have a $G$-site in
common.  Note that $L(G)$ is never bipartite, except for the trivial case
of a ring.  A well known example of a line graph is the kagome lattice.
(Incidentally, kagome is not a person---it is a pattern of woven bamboo.)

The hopping matrix $T$ is defined to be $-t < 0$ on every edge of
$\Lambda = L(G)$.  Not only is there a restriction on the magnitude of
$t_{xy}$ but we see, once again, that negative $t_{xy}$ is the natural
sign---as it is for Theorem 2.

Of course, as Mielke and Tasaki note, and as is also true in Theorem 2,
one can convert positive $t_{xy}$ into negative $t_{xy}$ by a
hole-particle transformation on both spin $\uparrow$ and spin
$\downarrow$.  This does not alter the interaction or the total spin,
$S$, but it changes $N$ into $N^\prime = 2 \vert \Lambda \vert - N$.
For positive, $t_{xy}$, then, we can transform to the negative $t_{xy}$
situation and conclude that $S = N^\prime /2$ when $N^\prime \leq
N_0$.  This translates into $S = \vert \Lambda \vert - N/2$ when $N
\geq 2 \vert \Lambda \vert - N_0$ (and not $\leq$).  This construction
would yield saturated ferromagnetism, $S = N/2$, only at $N = \vert
\Lambda \vert$ (half-filled band), provided it were possible to achieve
$N_0 \geq \vert \Lambda \vert$; this is clearly impossible (unless $T =
0$), so the positive $t_{xy}$ choice does {\it not\/} yield the desired
saturated ferromagnetism.  However, it is still possible to have
unsaturated ferromagnetism, i.e. $1 > 2S/N \not= 0$ in the
thermodynamic limit.  By taking $N = 2 \vert \Lambda \vert - N_0$ and
$S = \vert \Lambda \vert - N/2$ we have $2S/N = N_0/(2 \vert \Lambda
\vert - N_0)$. (One {\it can\/ } reasonably argue, however, that the
condition $S = \vert \Lambda \vert - N/2$ {\it is indeed\/} saturated
ferromagnetism because this value of the spin is the maximum possible
one when $N > \vert \Lambda\vert$, given the constraints of the sytem
and given the Pauli principle.  The system is constrained by allowing
only a limited number of states for the electrons, i.e., two per site.
I leave the semantic resolution to the reader.)

The homology of $G$ determines $\ch_0$ in a simple way.  Pick any
closed, self avoiding path in $G$ of even length.  This path
corresponds, in an obvious way, to a closed path, $P$, of the same
length in $L(G)$.  If, now, we take the vector $\phi (x) = 0$ for $x$
not a vertex of $P$ and $\phi (x) =$ alternately $+1$ and $-1$ as we
traverse the vertices of $P$, we see at once that $\phi$ is an
eigenvector with eigenvalue $-2t$.  It is also not hard to prove that
$-2t$ cannot be improved, i.e., $\lambda_0 = -2t$.  Moreover, this
construction yields all of $\ch_0$.

In [TH2, MT] essentially the same result as Mielke's (with similar
requirements on $T$) is achieved, but with certain decorated lattices
with next nearest neighbor hoppings.  The ferromagnetism of [TH2, MT]
is shown to be stable under small change of the electron density, by
using a ``grand canonical ensemble'' with a fixed electron density.  It
is also proved that there is a transition to a paramagnetic phase as
one decreases the electron density.  Thus (3.11) and (3.12) (with $S$
replaced by its grand canonical average) are proved for these special
graphs.

In all cases, the eigenvectors in $\ch_0$ can all be taken to have
compact support, i.e., each $\phi (x)$ vanishes except on a finite set
of sites of $\Lambda$, and each such set is independent of $\Lambda$
once $\Lambda$ is large enough to include the set.   This property
leads to the result [MT] that the effective Hamiltonian, in a suitable
subspace of states that includes the ground states, can be written
exactly as a Heisenberg Hamiltonian.
\bigskip
\bigskip
\noindent
{\bf 4. The Flux Phase Problem}
\medskip

Very little seems to be known rigorously about the effect of a magnetic
field on the orbital motion, that is, if we set
$$
t_{xy} = \vert t_{xy} \vert\,\exp [i \theta_{xy}], \ \qquad \theta_{xy} =
\int^y_x A\cdot
ds \eqno(4.1)$$
for some vector potential $A$.  The spin of the ground state might well
change.  The known proofs of Theorems 1 and 2 fail when $A \not= 0$.  If,
indeed,
the spin of the ground state changes then we have a new kind of ``magnetic
field---spin" interaction, brought about by the Pauli exclusion principle.
Indeed, something similar is discussed in the one-dimensional context in
[FK1].

The energy certainly does change, and it is by now well known that when $U =
0$ the zero flux state does {\it not} give the lowest ground state energy.
When $\Lambda$ is a square lattice and when $N = |\Lambda|$, it is
conjectured that the {\it maximum} (!) possible flux, namely $\pi$ in each
square
($t_{12} t_{23} t_{34} t_{41} = -1$ around a square), is optimum.  (For a
survey of this question and for some rigorous results about it see [LL2].)
The same conjecture has been made for $U \not= 0$.
\smallskip

{\bf Problem 5}: {\it Solve the flux phase problem for $U \not= 0$ (or even for
$U = 0$) for a half-filled band on a square lattice.}
\smallskip

The moral of this story is that the Pauli exclusion principle can really
upset our ideas about diamagnetism.  For one solitary electron, the
imposition of a magnetic field raises the ground state energy.  For many
electrons it can and does lower the energy.  It is trivially true, for
example, that when $U = +\infty$ and $N = |\Lambda|$, or $U = {\rm
anything}$ and $N = 2|\Lambda|$, a magnetic field has absolutely no effect on
the energy.
\vfill\eject
\noindent
{\bf 5. Uniform Density Theorem}
\medskip

The hole-particle transformation has remarkable consequences for the Hubbard
model on a bipartite graph with a half-filled band, $N = |\Lambda|$.  The
ideas given below are well known to chemists---less so to physicists.  They
go back to Coulson and Rushbrooke [CR] for the $U = 0 $ case, and to
MacLachlan [MA] who generalized them to many interacting
models---including the Hubbard model as a special case.  A simplified
proof together with an extension to models involving explicity spin-spin
interactions (such as the $t-J$ model), and to the Falicov-Kimball model,
is given in [LLM].

The results apply equally to three cases: (a) The canonical
Gibbs state with $N = |\Lambda|$; (b) The
grand-canonical Gibbs state with zero chemical potential;
(c) The ground state with $N =
|\Lambda|$ and which is defined, in case of degeneracy, to be the $\beta
\rightarrow \infty$ limit of the canonical Gibbs state.   The one-body
density matrix $\rho_\sigma (x,y)$ is the expectation value of
$c^\dagger_{x\sigma} c^{{\phantom{\dagger}}}_{y\sigma}$ in the state in
question.
\smallskip

{\bf Theorem 4}: {\it For a half-filled band on a bipartite lattice, the
one-body density matrix for each of the above states satisfies (for each
$\sigma = \uparrow {\rm or} \downarrow$)}.
$$
\rho_\sigma (x,y) = \mfr1/2 \delta_{xy} \qquad {\sl if}\ x,y \epsilon A \
{\sl or} \ x,y \epsilon B. \eqno(5.1)$$

If $x \epsilon A$ and $y \epsilon B$, nothing simple can be said.  Note
that the theorem does {\it not} require $T$ to be real, and is thus one of
the few theorems that applies to complex $t_{xy}$.  However, a true, physical
magnetic field would also act on the electron spins and thereby vitiate the
hole-particle symmetry needed for the proof.

The complex case is a bit subtle, for it uses more than just a hole-particle
transformation (call it $W$).  It also utilizes the nonlinear antiunitary map
$J$ that maps a vector $\psi$, considered as a polynomial in the
$c^\dagger_{x\sigma}$'s applied to the vacuum, into the vector $\psi^*$
corresponding to the polynomial with complex conjugate coefficients.
While $J$ is nonlinear, $JKJ$ is linear when K is any linear operator, and $Tr
JKJ = (TrK)^*$.  The antiunitary $Y = JW$ satisfies $Y = JW = WJ$ (and
hence
$Y^2 = \1$) and $Y c_{x\sigma} Y = W c_{x\sigma} W$.  Most important is the
invariance of the Hamiltonian, $YHY = H$, which replaces the hole particle
invariance, $WHW = H$, which fails for complex $T$.

What is the
significance of this result?  It seems to contradict the conjecture in
Problem 1 that there can be antiferromagnetic, i.e., staggered, LRO.  Such
an ordering can occur only in the thermodynamic limit and it has the
property that for every state with ordering $\uparrow \downarrow \uparrow
\downarrow \cdots$ there is a state with ordering $\downarrow \uparrow
\downarrow \uparrow \cdots$.  The point about Theorem 4, applied to the
thermodynamic limit, is that for every state with ordering there is an
equally good state (obtained by changing boundary conditions) with the
opposite ordering, and on the average each site will have the same density
for each spin value.  In other words, {\it there is no way to adjust the
potentials $U_x$ or the hopping matrix $t_{xy}$ in a clever way so as to
enhance the occupation of certain sites} --- in a manner independent of
boundary conditions.  This stability is remarkable and, although it is not
true for real materials, the theorem hints at some kind of remnant
stability that might transcend the overly idealized assumptions
needed for its proof.
\smallskip

{\bf Problem 6}: {\it Is there any residue of (5.1) when} $N \ne \vert
\Lambda \vert$?
\smallskip
\bigskip
\bigskip
\noindent
{\bf References}
\medskip

\item{[AL]}  M. Aizenman and E.H. Lieb, {\it Magnetic properties of some
itinerant-electron systems at $T>\nobreak0$},
Phys. Rev. Lett. {\bf 65}, 1470-1473 (1990).
\item{[AP]}  P.W. Anderson, {\it New approach to the theory of
superexchange interactions}, Phys. Rev. {\bf 115}, 2-13 (1959).
\item{[BH]}  H.A. Bethe, Zeits. f. Phys. {\bf 71}, 205-226 (1931); English
trans.: D.C. Mattis,
{\sl The Many-Body Problem},
World Scientific (1993), pp. 689-716.
\item{[CC]}  C.F. Coll III, {\it Excitation spectrum of the one-dimensional
Hubbard model}, Phys. Rev. B {\bf 9}, 2150-2159 (1974).
\item{[CCK]} S. Chakravarty, L. Chayes and S.A. Kivelson, {\it Absence of
pair binding in the $U = \infty$ Hubbard Model}, Lett. Math. Phys. {\bf
23}, 265-270 (1991).
\item{[CR]}  C.A. Coulson and G.S. Rushbrooke, {\it Note on the method
of molecular orbitals}, Proc. Cambridge Philos. Soc. {\bf 36}, 193-200
 (1940).
\item{[DLS]}  F.J. Dyson, E.H. Lieb and B. Simon, {\it Phase transitions in
quantum spin systems with isotropic and nonisotropic interactions}
J. Stat. Phys. {\bf 18}, 335-383 (1978).
\item{[DW]} B. Doucot and X.G. Wen, {\it Instability of the Nagaoka state
with more than one hole}, Phys. Rev. B {\bf 40}, 2719-2722 (1989).
\item{[EKS]}  F.H.L. Essler, V.E. Korepin and K. Schoutens, {\it Complete
solution of the one-dimensional Hubbard model}, Phys. Rev. Lett. {\bf 67},
3848-3851 (1991).  The details are in {\it Completeness of the $SO(4)$
extended Bethe ansatz for the one-dimensional Hubbard model}, Nucl. Phys. B
{\bf 384}, 431-458 (1982).
\item{[ES]}  F.H.L. Essler and V.E. Korepin, {\it Scattering matrix and
excitation spectrum of the Hubbard model}, preprint (1993).
\item{[FK1]}  S. Fujimoto and N. Kawakami, {\it Persistent currents in
mesoscopic Hubbard rings with spin-orbit interaction}, Yukawa Institute
preprint (July 1993).
\item{[FK2]}  L.M. Falicov and J.C. Kimball, {\it Simple model for
semiconductor-metal transitions:  $SmB_6$ and transition metal oxides},
Phys. Rev. Lett. {\bf 22}, 997-999 (1969).
\item{[FL1]}  M. Flicker and E.H. Lieb, {\it Delta function fermi gas with
two-spin deviates}, Phys. Rev. {\bf 161}, 179-188 (1967).
\item{[FL2]} J.K. Freericks and E.H. Lieb, {\it The ground state of a
general electron-phonon Hamiltonian is a spin singlet}, in preparation.
\item{[FRDS]} Y. Fand, A.E. Ruckenstein, E. Dagatto and S. Schmitt-Rink,
{\it Holes in the infinite U Hubbard model: Instability of the Nagaoka
state}, Phys. Rev. B {\bf 40}, 7406-7409 (1989).
\item{[GD]}  D.K. Ghosh, {\it Nonexistence of magnetic ordering in the
one-- and two--dimensional Hubbard model}, Phys. Rev. Lett. {\bf 27},
1584-1586 (1971), [Errata, {\bf 28} 330 (1972)].
\item{[GH]}  H. Grosse, {\it The symmetry of the Hubbard model}, Lett. Math.
Phys. {\bf 18}, 151-156 (1989).
\item{[GM]}  M. Gaudin, {\it Un syst\`eme \`a une dimension de fermions en
interaction}, Phys. Letters {\bf 24A},  55-56 (1967).
\item{[GMC]}  M.C. Gutzwiller, {\it The effect of correlation on the
ferromagnetism of transition metals}, Phys. Rev. Lett. {\bf 10}, 159-162
(1963).
\item{[HJ]}  J. Hubbard, {\it Electron correlations in narrow energy
bands}, Proc. Roy. Soc. (London), {\bf A276}, 238-257
(1963).
\item{[HL]}  O.J. Heilmann and E.H. Lieb, {\it Violation of the
non-crossing rule: the Hubbard Hamiltonian for benzene}, Trans. N.Y.
Acad.  Sci. {\bf 33}, 116-149 (1970).  Also in Ann. N.Y. Acad. Sci.
{\bf 172}, 583-617 (1971).
\item{[HM]} Th. Hanisch and E. M\"uller-Hartmann: {\it Ferromagnetism in the
Hubbard Model:  Instability of the Nagaoka State on the Square
Lattice}, Ann. Physik {\bf 2}, 381-397 (1993); See also
E. M\"uller-Hartmann, Th. Hanisch and R. Hirsch: {\it Ferromagnetism of Hubbard
Models}, Physica B {\bf 186-188}, 834-836 (1993).
\item{[KJ]}  J. Kanamori, {\it Electron correlation and ferromagnetism of
transition metals}, Prog. Theor. Phys. {\bf 30}, 275-289 (1963).
\item{[KK]} K. Kubo and K. Kishi, {\it Rigorous bounds on the
susceptibility  of the Hubbard model}, Phys. Rev. B {\bf 41}, 4866-4868
(1990).
\item{[KL]}  T. Kennedy and E.H. Lieb, {\it An itinerant electron model
with crystalline or magnetic long range order}, Physica {\bf 138A}, 320-358
(1986).
\item{[KLS]} T. Kennedy, E.H. Lieb and S. Shastry, {\it Existence of
N\'eel order in some spin 1/2 Heisenberg antiferromagnets}, J. Stat. Phys. {\bf
53}, 1019-1030 (1988).
\item{[KO]}  T. Koma, {\it An extension of the thermal Bethe ansatz --
one-dimensional Hubbard model}, Prog. Theor. Phys. {\bf 83}, 655-659
(1990).
\item{[KT]}  T. Koma and H. Tasaki, {\it Decay of superconducting and
magnetic correlations in one- and two-dimensional Hubbard models},
Phys. Rev. Lett. {\bf 68}, 3248-3251 (1992).
\item{[LE1]}  E.H. Lieb, {\it Two theorems on the Hubbard model}, Phys.
Rev. Lett. {\bf 62}, 1201-1204 (1989), [Errata {\bf 62}, 1927 (1989)].
\item{[LE2]} E.H. Lieb, {\it Models}, in Proceedings of the Solvay
institute 14th conference on chemistry at the University of Brussels,
May 1969, Phase transitions, Interscience, 1971.
\item{[LL1]}  E.H. Lieb and W. Liniger, {\it Exact analysis of an
interacting Bose gas. I. The general solution and the ground state},
Phys. Rev. {\bf 130}, 1605-1616 (1963).
\item{[LL2]} E.H. Lieb and M. Loss, {\it Fluxes, Laplacians and
Kasteleyn's theorem}, Duke Math. J. {\bf 71}, 337-363 (1993).
\item{[LLM]} E.H. Lieb, M. Loss and R.J. McCann, {\it Uniform density theorem
for the Hubbard model}, J. Math. Phys. {\bf 34}, 891-898 (1993).
\item{[LM1]}  E.H. Lieb and D.C. Mattis, {\it Theory of ferromagnetism and
the ordering of electronic energy levels}, Phys. Rev. {\bf 125}, 164-172
(1962).
\item{[LM2]}  E.H. Lieb and D.C. Mattis, {\it Ordering energy levels of
interacting spin systems}, J. Math. Phys. {\bf 3}, 749-751 (1962).
\item{[LvE]} W. von der Linden and D.M. Edwards, {\it Ferrromagnetism
in the Hubbard model}, J. Phys. Cond. Matt. {\bf 3}, 4917-4940 (1991).
\item{[LW]}  E.H. Lieb and F.Y. Wu, {\it Absence of Mott transition in an
exact solution of the short-range, one-band model in one dimension},
Phys. Rev. Lett. {\bf 20}, 1445-1448 (1968).
\item{[MA]}  A.D. MacLachlan, {\it The pairing of electronic states in
alternant hydrocarbons}, Mol.~Phys. {\bf 2}, 271-284 (1959); {\it
Electrons and holes in alternant hydrocarbons}, Mol.~Phys.~{\bf 4},
49-56 (1961).
\item{[Mi1]} A. Mielke, {\it Ferromagnetic ground states for the Hubbard
model on line graphs}, J. Phys. A {\bf 24}, L73-L77 (1991); {\it
Ferromagnetism in the Hubbard model on line graphs and further
considerations}, J. Phys. A {\bf 24}, 3311-3321 (1991); {\it Exact ground
states for the Hubbard model on the kagome lattice}, J. Phys. A {\bf 25},
4335-4345 (1992); {\it Ferromagnetism in the Hubbard model and Hund's
rule}, Phys. Lett. A {\bf 174}, 443-448 (1993).
\item{[Mi2]} A. Mielke, {\it The one-dimensional Hubbard model for large or
infinite U}, J. Stat. Phys. {\bf 62}, 509-528 (1991).
\item{[MJ]}  J.B. McGuire, {\it Interacting fermions in one dimension. I.
Repulsive potential}, J. Math. Phys. {\bf 6}, 432-439 (1965).
\item{[MS]}  O. McBryan and T. Spencer, {\it On the decay of correlations
in $SO (n)$-symmetric ferromagnets}, Commun. Math. Phys. {\bf 53}, 299-302
(1977).
\item{[MT]}  A. Mielke and H. Tasaki, {\it Ferromagnetism in the Hubbard
model}, Commun. Math. Phys. (in press).
\item{[MV]}  W. Metzner and D. Vollhardt, {\it Correlated lattice fermions in
$d = \infty$ dimensions}, Phys. Rev. Lett. {\bf 62}, 324-327 (1989).
\item{[NY]} Y. Nagaoka, {\it Ferromagnetism in a narrow, almost
half-filled s band}, Phys. Rev. {\bf 147}, 392-405 (1966).
\item{[OA]}  A.A. Ovchinnikov, Zh.Eksp. Teor. Fiz. {\bf 57}, 2137-2143
(rd-review.tex1969).
Engl. trans. {\it Excitation spectrum in the one-dimensional
Hubbard model}, Sov. Phys. JETP {\bf 30}, 1160-1163 (1970).
\item{[PJ]}  J.A. Pople, {\it Electron interaction in unsaturated
hydrocarbons}, Trans. Faraday Soc. {\bf 49},  1375-1385 (1953).
\item{[PP]}  R. Pariser and R.G. Parr, {\it A semi-empirical theory of
the electronic spectra and electronic structure of complex unsaturated
hydrocarbons I. and II.}, J. Chem. Phys. {\bf 21}, 466-471, 767-776 (1953).
\item{[SA1]} A. S\"ut\H o, {\it Absence of highest spin ground states in the
Hubbard model}, Commun. Math. Phys. {\bf 140}, 43-62 (1991).
\item{[SA2]}  A. S\"ut\H o, {\it Bounds on ferromagnetism at $T > 0$ in the
hard-core lattice model}, Phys. Rev. B. {\bf 43}, 8779-8781 (1991); {\it
The $U = \infty$ Hubbard model at positive temperature} in
{\sl From Phase Transitions to Chaos}, G. Gyorgi ed., World Scientific (1992).
\item{[SB]}  B.S. Shastry, {\it Infinite conservation laws in the
one-dimensional Hubbard model}, Phys. Rev. Lett. {\bf 56}, 1529-1531 (1986),
[Errata {\bf 56}, 2334 (1986)] and {\it Exact integrability of the
one-dimensional Hubbard model}, Phys. Rev. Lett. {\bf 56}, 2453-2456
(1986).  The general method is clarified in {\it Decorated star-triangle
relations and exact integrability of the one-dimensional Hubbard model}, J.
Stat. Phys. {\bf 50}, 57-79 (1988).
\item{[SH]}  H. Shiba, {\it Magnetic susceptibility at zero temperature for
the one-dimensional Hubbard model}, Phys. Rev. B {\bf 6}, 930-938 (1972).
\item{[SKA]} B.S. Shastry, H.R. Krishnamurthy and P.W. Anderson, {\it
Instability of the Nagaoka ferromagnetic state of the $U=\infty$ Hubbard
model}, Phys. Rev. B {\bf 41}, 275-2379 (1990).
\item{[SQT]} S.Q. Shen, Z.M. Qiu and G.S. Tian, {\it The Nagaoka state and
its stability in the one-band Hubbard model}, Phys. Lett. A {\bf 178},
426-430 (1993).
\item{[TB]}  B. T\'oth, {\it Failure of saturated ferromagnetism for the
Hubbard model with two holes}, Lett. Math. Phys. {\bf 22}, 321-333 (1991).
\item{[TD]} D.J. Thouless, Proc. Phys. Soc. (London), {\it Exchange in
Solid $^3He$ and the Heisenberg Hamiltonian} {\bf 86}, 893-904 (1965).
\item{[TG1]} G.S. Tian, {\it A simplified proof of Nagaoka's theorem}, J.
Phys. A {\bf 23}, 2231-2236 (1990).
\item{[TG2]} G.S. Tian, {\it The Nagaoka state in the one-band Hubbard model
with two and more holes}, J. Phys. A {\bf 24} 513-521 (1991).
\item{[TG3]} G.S. Tian, {\it Stability of the Nagaoka state in the one-band
Hubbard model}, Phys. Rev. B {\bf 44}, 4444-4448 (1991).
\item{[TH1]} H. Tasaki, {\it Extension of Nagaoka's theorem on the large
$U$ Hubbard Model}, Phys. Rev. B {\bf 40}, 9192-9193 (1989).
\item{[TH2]}  H. Tasaki, {\it Ferromagnetism in Hubbard models with
degenerate single-electron ground states}, Phys. Rev. Lett. {\bf 69},
1608-1611 (1992).
\item{[TM]}  M. Takahashi, {\it Magnetization curve for the half-filled
Hubbard model}, Prog. Theor. Phys. {\bf 42}, 1098-1105 (1969) and {\it
Magnetic susceptibility for the half-filled Hubbard model},
Prog. Theor. Phys. {\bf 43}, 1619 (1970).
\item{[TS]}  S.A. Trugman, {\it Exact results for the $U = \infty$ Hubbard
model}, Phys. Rev. B {\bf 42}, 6612-6613 (1990).
\item{[UTS]}  K. Ueda, H. Tsunetsugu and M. Sigrist, {\it Singlet ground
state of the periodic Anderson model at half filling:  a rigorous result},
Phys. Rev. Lett. {\bf 68}, 1030-1033 (1992).
\item{[WE]}  F. Woynarovich and H.P. Eckle, {\it Finite size corrections
for the low lying states of a half-filled Hubbard chain},
J. Phys. A {\bf 20},  L443-449 (1987).
\item{[WF]}  F. Woynarovich, {\it Excitations with complex wavenumbers in a
Hubbard chain:  I. States with one pair of complex wavenumbers}, J. Phys. C
{\bf 15}, 85-96 and {\it II.  States with several pairs of complex
wavenumbers}, 97-109 (1982).
\item{[WR]} M.B. Walker and Th.W. Ruijgrok, {\it Absence of magnetic
ordering in one and two dimensions in a many-band model for interacting
electrons in a metal}, Phys. Rev. {\bf 171}, 513-515 (1968).
\item{[YC]}  C.N. Yang, {\it Some exact results for the many-body problem
in one-dimension with repulsive delta-function interaction}, Phys. Rev.
Lett. {\bf 19}, 1312-1315 (1967).
\item{[YZ]} C.N. Yang and S.C. Zhang, {\it $SO_4$ symmetry in a Hubbard
model}, Mod. Phys. Lett. {\bf B4}, 759-766 (1990).


\bye